\documentstyle[referee,epsf]{mn}
\input psfig.tex
\def\lsim{\lower.5ex\hbox{$\; \buildrel < \over \sim \;$}}
\def\gsim{\lower.5ex\hbox{$\; \buildrel > \over \sim \;$}}
\newif\ifAMStwofonts
\ifoldfss
  \ifCUPmtlplainloaded \else
    \NewTextAlphabet{textbfit} {cmbxti10} {}
    \NewTextAlphabet{textbfss} {cmssbx10} {}
    \NewMathAlphabet{mathbfit} {cmbxti10} {} 
    \NewMathAlphabet{mathbfss} {cmssbx10} {} 
  \fi
  \ifAMStwofonts
    \ifCUPmtlplainloaded \else
      \NewSymbolFont{upmath} {eurm10}
      \NewSymbolFont{AMSa} {msam10}
      \NewMathSymbol{\upi}     {0}{upmath}{19}
      \NewMathSymbol{\umu}     {0}{upmath}{16}
      \NewMathSymbol{\upartial}{0}{upmath}{40}
      \NewMathSymbol{\leqslant}{3}{AMSa}{36}
      \NewMathSymbol{\geqslant}{3}{AMSa}{3E}

    \fi
  \fi
\fi 
\ifnfssone
  \newmathalphabet{\mathit}
  \addtoversion{normal}{\mathit}{cmr}{m}{it}
  \addtoversion{bold}{\mathit}{cmr}{bx}{it}
  \newmathalphabet{\mathbfit} 
  \addtoversion{normal}{\mathbfit}{cmr}{bx}{it}
  \addtoversion{bold}{\mathbfit}{cmr}{bx}{it}
  \newmathalphabet{\mathbfss} 
  \addtoversion{normal}{\mathbfss}{cmss}{bx}{n}
  \addtoversion{bold}{\mathbfss}{cmss}{bx}{n}
  \ifAMStwofonts
    \ifCUPmtlplainloaded \else
      \UseAMStwoboldmath
      \makeatletter
      \new@mathgroup\upmath@group
      \define@mathgroup\mv@normal\upmath@group{eur}{m}{n}
      \define@mathgroup\mv@bold\upmath@group{eur}{b}{n}
      \edef\UPM{\hexnumber\upmath@group}
      \new@mathgroup\amsa@group
      \define@mathgroup\mv@normal\amsa@group{msa}{m}{n}
      \define@mathgroup\mv@bold\amsa@group{msa}{m}{n}
      \edef\AMSa{\hexnumber\amsa@group}
      \makeatother
      \mathchardef\upi="0\UPM19
      \mathchardef\umu="0\UPM16
      \mathchardef\upartial="0\UPM40
      \mathchardef\leqslant="3\AMSa36
      \mathchardef\geqslant="3\AMSa3E
    \fi
  \fi
\fi 

\ifnfsstwo
  \DeclareMathAlphabet{\mathbfit}{OT1}{cmr}{bx}{it}
  \SetMathAlphabet\mathbfit{bold}{OT1}{cmr}{bx}{it}
  \DeclareMathAlphabet{\mathbfss}{OT1}{cmss}{bx}{n}
  \SetMathAlphabet\mathbfss{bold}{OT1}{cmss}{bx}{n}
  \ifAMStwofonts
    \ifCUPmtlplainloaded \else
      \SetSymbolFont{UPM}{bold}{U}{eur}{b}{n}
      \DeclareSymbolFont{AMSa}{U}{msa}{m}{n}
      \DeclareMathSymbol{\upi}{0}{UPM}{"19}
      \DeclareMathSymbol{\umu}{0}{UPM}{"16}
      \DeclareMathSymbol{\upartial}{0}{UPM}{"40}
      \DeclareMathSymbol{\leqslant}{3}{AMSa}{"36}
      \DeclareMathSymbol{\geqslant}{3}{AMSa}{"3E}
    \fi
  \fi
\fi 

\ifCUPmtlplainloaded \else
  \ifAMStwofonts \else 
    \def\upi{\pi}
    \def\umu{\mu}
    \def\upartial{\partial}
  \fi
\fi

\title{X-ray Observation of SS 433 with RXTE} 

\author[A. Nandi, Sandip K. Chakrabarti, Tomaso Belloni and P. Goldoni]
{A. Nandi$^1$, Sandip K.\ Chakrabarti$^{1,2}$, Tomaso Belloni$^{3}$ and P. Goldoni$^{4}$\\
$^1$S.N. Bose National Center for Basic Sciences,\\
JD-Block, Salt Lake, Kolkata, 700098, India\\
$^2$ Centre for Space Physics, Chalantika 43, Garia Station Rd., 700084, India\\
$^3$ INAF-Osservatorio Astronomico di Brera, Mirate, Italy\\ 
$^4$ CEA Saclay, Service D' Astrophysique, France\\ }

\begin{document}

\maketitle

\begin{abstract}
Apart from regular monitoring by ASM, the compact object SS 433 was observed with RXTE several times
last two/three years. We present the first analysis of these observations.  We also include 
the results of the recent exciting TOO campaign made during donour inferior 
(orbital phase $\phi=0$) and superior 
($\phi=0.5$) conjunctions which took place on Oct. 2nd, 2003, and on March 13th, 2004 respectively, 
when the jet itself was directly pointing towards us (i.e.,  precessional phase $\psi \sim 0$). 
Generally, we found that two distinct lines fit the spectra taken on all these days. 
We present some of the light-curves and the X-ray spectra, and show that the Doppler 
shifts of the emitted lines roughly match those predicted by the kinematic model
for the jets. We find that the line with a higher energy can be best identified with a FeXXVI
Ly-$\alpha$ transition while the line with lower energy can be identified with a FeXXV 
(1s2p - 1s$^2$) transition. We observe that the X-ray flux on March 13th, 2004 (when the base 
of the jet is exposed) is more than twice compared to that on Oct. 2nd, 2003
(when the base is covered by the companion). We find the flux to continue to remain high
at least till another orbital period. We believe that this is because SS 433 was undergoing
a weak flaring activity during the recent observation. 
\end{abstract}

\noindent SUBMITTED APRIL, 2004; REVISED OCTOBER, 2004 (SUBMITTED TO MNRAS)

\section{Introduction}

The galactic micro-quasar SS 433 is an enigmatic X-ray binary system,
ejecting material in the form of bipolar jets with almost
constant velocity ($v \sim 0.26c$) (Margon 1984; Gies et al. 2002).
Even after $25$ years of its discovery, it is not clear whether the
compact object is a neutron star or a black hole. The nature of the
companion is also not known with certainly. Recently, Gies et al. (2002)
pointed out from their UV spectroscopic study of the mass
donour star of SS 433 that the nature of the companion is probably an evolved
A-type star with a black hole as the primary. The disk, along with the jet which
is along the instantaneous normal to it, itself precesses with a $162.15$ day periodicity.
The binary has a $13.1$ day periodicity (Margon 1984). There are observational supports
in optical as well as in radio/IR/X-rays (Borisov \& Fabrika, 1987; Vermeulen et al. 1993ab;
Chakrabarti et al. 2002, 2003; Migliari, Fender and M\'endez, 2002)
support that the jets are ejected in a bullet-like fashion,
i.e., as blobs ejected along the instantaneous axis of the disk having only
the {\it radial velocity component}. The red/blue shifts of the lines emitted
from these jets are described very accurately by the so-called kinematic
model of Abell \& Margon (1979). These blobs in Radio/X-ray are seen
within a few arcseconds from the core, although radio emissions and X-rays are also present
much farther out, in the scale of half a degree (e.g., Margon, 1984; Migliari, Fender \& Mend\'ez, 2002).

Although extensive optical and radio monitoring of the source has provided the
basic parameters to describe the disk-jet system, there are very few
observation in X-rays. In X-rays, SS 433 is a relatively weak source
and is not generally observable in hard X-rays beyond $\sim 30$ keV.
Previous observations by EXOSAT, Ginga, ASCA confirmed the existence 
of the Doppler-shifted X-ray emission lines (Watson et al. 1986; Yuan et al. 1995;
Kotani et al. 1994). The shifting of the Fe-line
was found to be consistent with the predictions of the kinematic model. This
therefore indicated that the X-ray emitting material is physically
associated with the jets. Recently, the Chandra observation of SS 433 spectrum
(Marshall et al. 2002, Namiki et al. 2003) shows very much complex behaviour with a large number
of blue and red shifted lines. Kinematic model was also established, but higher
velocity was required for the line emitting gases. 
Iron line emission from the extended region of the jet has also been observed
by Chandra (Migliari, Fender and M\'endez, 2002).

In the present paper, we report results obtained using the RXTE satellite, compiling
archival data as well as TOO data triggered by us. Most importantly, for the
first time, the spectrum obtained during (a) the {\it inferior conjunction}
(when the central compact object and the base of the jet 
is blocked exactly by the companion; i.e., the orbital phase $\phi=0$)
{\it and} at the same time, when the precessional phase, $\psi$ was also $0$
and (b) the {\it superior conjunction} (when the companion is hidden by the disk 
and the jet is completely exposed to us, i.e., the orbital phase $\phi=0.5$)
{\it and} at the same time, when the precessional phase, $\psi$ was also $0$
are presented and analyzed. In these special days, the jet was directly pointing
towards us emitting lines with the highest possible line shifts.
These data were obtained on Oct. 2nd, 2003 and on Mar. 13th, 2004. Because of lower energy
resolution of RXTE/PCA instrument ($<18$\% at $6$keV), it is not often easy to identify the
exact source of the lines and line flux would also be inaccurate due to blending with other lines.
Fortunately comparing results with the earlier observations by  ASCA (Kotani et al. 1996) and Chandra 
(Marshall et al. 2002, Namiki et al. 2003) the identification became easier. 

In the next section, we present the Tables containing a log of observations analysed by us and the results
of our analysis. We also present the
light curves of some of these observations. In Section 3, we present fits of the spectrum
obtained by us at three epochs, including the ones at the inferior and the superior conjunctions. We also
show how the line energy (including red/blue shifts) match with the shifts predicted from the kinematic
model. In Section 4, we discuss the possibility of X-ray flares in this system.
Finally, in Section 5, we draw our conclusions.

\section {Observations}

SS 433 was pointed at several times by Proportional Counter Array (PCA) detectors 
on board RXTE satellite. The PCA contains five ($0$ $...$ $4$)  Proportional 
Counter Units (PCUs). We concentrate on observations taken since November, 
2001. In order to avoid biasing the analysis, we selected only those observations which 
were taken by the same units of PCA detector, namely, $2$ \& $3$. We made the
analysis by adding data from these two PCUs together. A log of these
observations is given in Table 1. The first Column gives the log of 
observation and the second column gives the date and time when the
observation begins (in MJD) along with the RXTE Observation ID.
To calculate the precessional phase ($\psi$) and orbital phase ($\phi$), at the begining
time of these observations, we adopt the following ephemeris (Goranskii et al. 1998):
HJD $2451458.12+162.15$E for $\psi$ and HJD $2450023.62+13.0821$E for $\phi$.
These are given in Column 3. Observations I (made on Oct. 2nd, 2003) and observation K 
(made on Mar. 13th, 2004) had $\phi\sim 0.0$ and $\phi\sim 0.5$ respectively
both having  $\psi\sim 0$. TOO observations G-L were triggered at our request.  A 
recent TOO data (Observation M made on 25th of March, 2004) is also included. 
We also re-analysed the archival data of Nov. 2001. The spectra were fitted
with a thermal bremsstrahlung (TB) and iron line(s) and the temperature is given in Column 4
(See next section for details of the fitting procedure.)
The total integrated flux of X-rays in $3-25$keV range in units of $10^{-10}$ergs cm$^{-2}$ sec$^{-1}$ is
presented in Column 5. During the recent superior conjunction when the jet was exposed to us (March 13th, 2004),  
the net flux was found to be more than twice as high compared to that in the inferior conjunction
(Oct. 2nd, 2003). The net flux remained high for at least another orbital period indicating that 
SS 433 may be undergoing a weak flaring activity.

One can compare the X-ray flux by RXTE observations with earlier results of EXOSAT (Watson et al. 1986)
and Chandra (Namiki et al. 2003). For instance, in the same unit as in Table 1, at $\psi\sim 0$, 
the 2-6keV flux of EXOSAT was $1.5$ while our fluxes vary from $1.07-1.6$ depending on $\phi$ in the same 
range of 2-6keV. Chandra, on the other hand, found the total integrated flux in 1-10keV at $\psi\sim 0.4$
to be $0.9$.  We did not make any observation at this phase, and  
our result at phase $\psi\sim 0$ in $3-9$keV is $1.6-3$. Thus the fluxes measured by
RXTE are comparable to previous measurements, although an actual comparison requires
observation to be carried out at the same precessional and orbital phases as well.

According to the kinematic model (Abell \& Margon, 1979; Margon 1984), the red/blue 
shifts $z$ of the emitted line are computed from:
$$
1+z = \gamma(1 \pm v_j ~ {\rm sin} \theta ~ {\rm sin} i ~ {\rm cos} \psi \pm v_j ~
{\rm cos} \theta ~ {\rm cos} i),
\eqno{(1)}
$$
where, $v_j$ (taken here to be $0.2602$, see, Gies et al. 2002) is the proper velocity 
of the line emission in units of velocity of light $c$ from the jet matter, $i=78^o.83$
is the angle that the normal to the disk subtends with the line of sight,
$\theta=19^o.85$ is the angle subtended by the jet with the disk-normal, $\psi$
is the precession phase taken from Table 1. In this convention, a negative $z$
corresponds to blue-shifts and a positive $z$ corresponds to red-shifts. Note that we use the
definition of $\psi$ such that $\psi=0$ when the shifts of the blue and red-jets are
maximally different as in Margon (1984). 
In Table 2, we presented these values in Column 2. The significance of the
Columns $3$-$10$ will be discussed in the next Section.

\begin{table}
\scriptsize
\centering
\caption{RXTE Observation log$^{a}$}
\vskip 0.5cm
\begin{tabular}{|c|c|c|c|c|}
\hline
Obs.&MJD(UT) & $\psi$ & $kT_{b}$ & Flux$^b$  \\
log & (Date) &  &  &   \\
&ObsID   & $\phi$ &  (keV)  &  \\ \hline

A&52222(07:10:27) &0.716 & 23.02 & 2.824   \\ 
& (09th Nov'01) & &  &   \\ 
&60058-01-01-00 &0.106& $\tiny^{+1.19}_{-1.10}$ & \\ \hline

B&52224(06:47:29) & 0.728 & 27.03 &  2.966  \\ 
& (11th Nov'01) & &  &   \\ 
& 60058-01-03-00 & 0.257 & $\tiny^{+1.85}_{-1.56}$ & \\ \hline

C&52227(06:11:11) &0.747& 26.91 &  3.436 \\ 
& (14th Nov'01) & &  &   \\ 
&60058-01-06-00& 0.485 & $\tiny^{+1.62}_{-1.41}$ & \\ \hline

D&52228(06:00:16) &0.753 & 24.01 &  3.480   \\ 
& (15th Nov'01) & &  &   \\ 
&60058-01-07-00& 0.561& $\tiny^{+1.45}_{-1.19}$ & \\ \hline

E&52234(08:03:15) &0.790& 25.93 & 2.786 \\ 
& (21st Nov'01) & &  &   \\ 
&60058-01-12-00& 0.025& $\tiny^{+7.53}_{-4.44}$ & \\ \hline

F&52235(07:54:37)$\bf^c$ &0.796& 27.63 & 3.547  \\ 
& (22nd Nov'01) & &  &   \\ 
&60058-01-13-00& 0.102& $\tiny^{+1.83}_{-1.54}$ & \\ \hline

G&52545(15:57:34) &0.705& 13.92 & 2.375   \\ 
& (27th Sep'02) & &  &   \\ 
&70416-01-01-01& 0.747& $\tiny^{+0.57}_{-0.71}$ & \\ \hline

H&52914(16:41:24) &0.981& 17.23 & 3.183 \\ 
& (01st Oct'03) & &  &   \\ 
&80429-01-01-00& 0.956& $\tiny^{+0.71}_{-0.65}$ & \\ \hline

I&52914(05:20:00)$\bf ^d$ &0.984& 17.50 & 2.993  \\ 
& (02nd Oct'03) & &  &   \\ 
&80429-01-01-01& 0.997& $\tiny^{+0.91}_{-0.81}$  & \\ \hline

J&53077(18:55:12) & 0.986 & 51.35 &  7.557 \\ 
& (12th Mar'04) & &  &   \\ 
&90401-01-01-01& 0.423& $\tiny^{+5.00}_{-3.18}$ & \\ \hline

K&53078(18:33:04)$\bf ^e$ &0.992& 44.93 & 6.912  \\ 
& (13th Mar'04) & &  &   \\ 
&90401-01-01-00& 0.498& $\tiny^{+2.60}_{-2.19}$ & \\ \hline

L&53079(18:12:00) &0.998 & 40.10& 6.778  \\
& (14th Mar'04) & &  &   \\ 
&90401-01-01-02& 0.574& $\tiny^{+1.99}_{-1.72}$ & \\ \hline

M&53089(01:51:28) &0.062 & 46.41 & 7.592  \\ 
& (25th Mar'04) & &  &   \\ 
&90401-01-02-01& 0.363& $\tiny^{+4.17}_{-3.40}$ & \\ \hline

\end{tabular}

\noindent{\small $a)$ Error bars in temperature are at $90$\% confidence level;
$b)$ Flux is in the range ($3-25$keV) in units of
$10^{-10}$ergs cm$^{-2}$ sec$^{-1}$; $c)$  Massive 
radio flare observed (Safi-Harb \& Kotani, 2003); \  $d)$ TOO observation 
at $\psi\sim 0$ and $\phi\sim 0$; \ $e)$ TOO 
observation at $\psi\sim 0$ and $\phi\sim 0.5$}\\
\end{table}


\begin {figure}
\vbox{
\centerline{
\psfig{figure=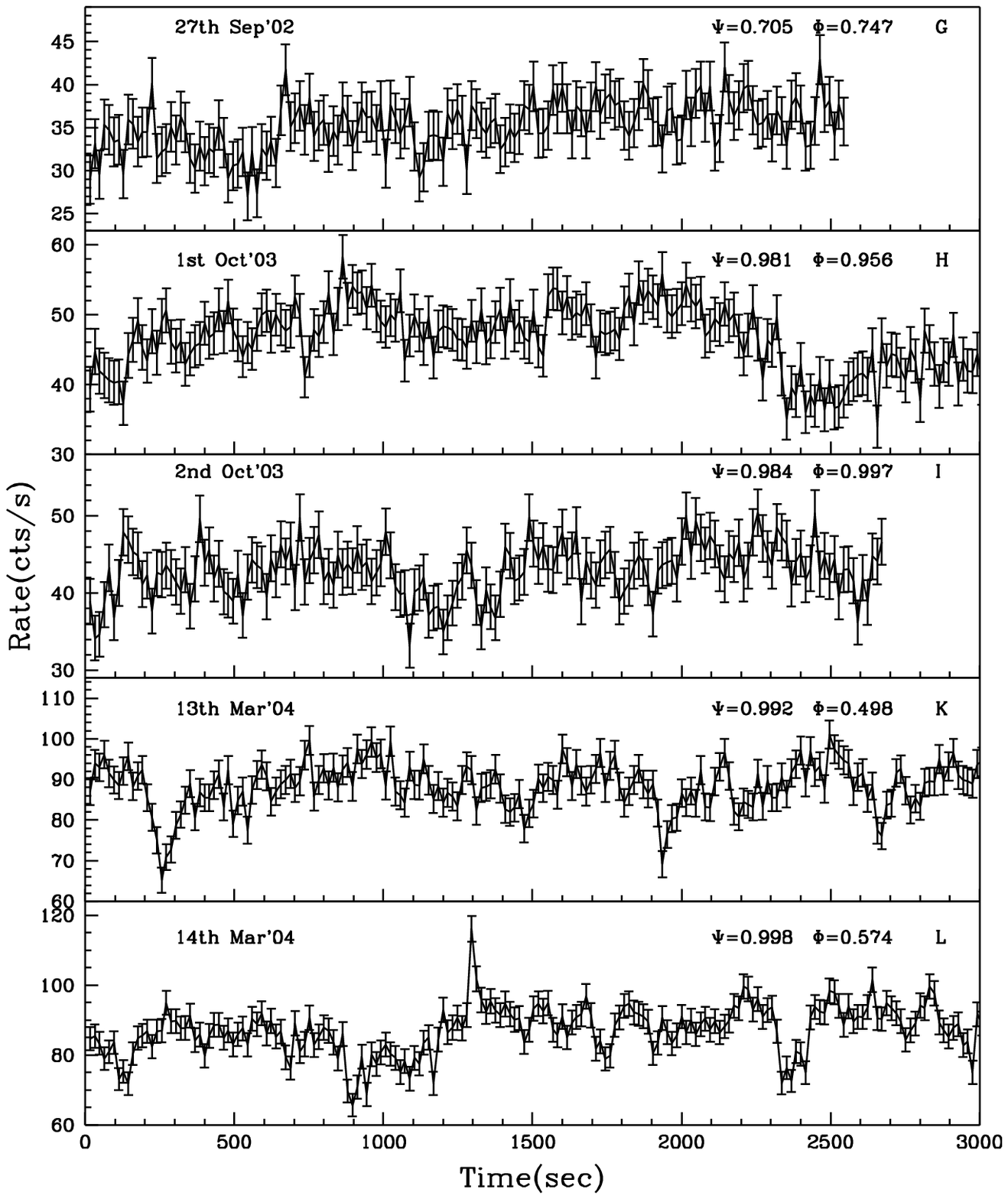,height=10truecm,width=14truecm}}}
\noindent{\small {\bf Fig. 1:} 
X-ray light curves (background subtracted) of SS433 extracted from $\bf Standard2$ data
for observations G , H , I, K and L (marked).  Along X-axis is time (in seconds) and along Y-axis is the count rate.}
\end{figure}

\begin{table}
\scriptsize
\centering
\caption{Results$^a$ of RXTE observations of SS 433}
\vskip 0.5cm
\begin{tabular}{|c|c|c|c|c|c|c|c|c|c|}
\hline
Obs.& $z_{blue}$ &  $\chi_1^2$ & $\chi_2^2$ &  $P_{F-stat}$ & $E_{obs}$ & $\sigma_{line}$&$F_{line}^b$ & z & z  \\
log& $z_{red}$ & ($\chi^2$/dof) & ($\chi^2$/dof) &  & (keV) & (keV) &($\times10^-3$) & FeXXVI & FeXXV \\ \hline

A& 0.005 & 0.742 & 0.633 & $1.7\times10^{-2}$ & $6.822 \tiny^{+0.038}_{-0.039}$ & $ 0.692\tiny^{+0.054}_{-0.053}$ & $ 2.536\tiny^{+0.144}_{-0.135}$ &\bf 0.020 $\tiny^{-0.005}_{+0.005}$ &-0.021 \\  \cline{9-10}
& 0.066 & (35.6/48) & (28.5/45) & ($2.1\sigma$) & -- & -- & -- & --  & -- \\ \hline

B&-0.001 & 0.966 & 0.682 & $4.9\times10^{-4}$ & $6.905\tiny^{+0.138}_{-0.153}$ & $1.050\tiny^{+0.209}_{-0.169}$ & $2.026\tiny^{+0.269}_{-0.271}$ &\bf 0.009 $\tiny^{-0.019}_{+0.021}$ &-0.033  \\ \cline{9-10}
& 0.073 & (46.4/48) & (31.4/46)& ($2.8\sigma$) & $6.819\tiny^{+0.002}_{-0.001}$ & 0.309$\pm${\tiny 0.0}& $0.808\tiny^{+0.330}_{-0.460}$ &  0.021 &\bf -0.020 $\tiny^{-0.0003}_{+0.0002}$  \\ \hline

C&-0.012 & 1.261 & 0.775 & $5.1\times10^{-6}$ & $6.976\tiny^{+0.081}_{-0.088}$ & 0.125$\pm${\tiny 0.0}& $0.613\tiny^{+0.164}_{-0.179}$ &\bf-0.002 $\tiny^{-0.012}_{+0.013}$ &-0.044 \\ \cline{9-10}
& 0.083 & (60.5/48)& (35.6/46) & ($3.5\sigma$) & $6.891\tiny^{+0.084}_{-0.099}$ & $1.159\tiny^{+0.132}_{-0.114}$ & $3.216\tiny^{+0.262}_{-0.246}$ &  0.011 &\bf -0.031 $\tiny^{-0.012}_{+0.015}$  \\ \hline

D&-0.015 & 1.819 & 1.265 & $3.6\times10^{-4}$ & $6.986\tiny^{+0.067}_{-0.074}$& 0.246$\pm${\tiny0.0} & $0.894\tiny^{+0.184}_{-0.213}$ &\bf-0.003 $\tiny^{-0.009}_{+0.011}$ &-0.045  \\ \cline{9-10}
& 0.087 & (87.3/48) & (58.2/46)& ($2.9\sigma$) & $6.837\tiny^{+0.128}_{-0.170}$ & $1.209\tiny^{+0.198}_{-0.160}$ & $2.632\tiny^{+0.305}_{-0.271}$ &  0.018 &\bf -0.023 $\tiny^{-0.019}_{+0.025}$  \\ \hline

E$^c$&-0.036 & 1.373 & 1.027& $1.7\times10^{-3}$ & $7.281\tiny^{+0.136}_{-0.154}$ & $1.227\tiny^{+0.184}_{-0.171}$ & $3.212\tiny^{+0.745}_{-0.595}$ &\bf-0.045 $\tiny^{-0.019}_{+0.022}$ & -0.089 \\ \cline{9-10}
& 0.107 & (65.9/48)& (47.2/46) & ($2.6\sigma$) &--&-- &--& -- & --  \\ \hline

F$\bf^d$ &-0.039 & 1.011 & 0.745 & $3.3\times10^{-4}$ & $6.950\tiny^{+0.092}_{-0.104}$ & 0.257$\pm${\tiny 0.0} & $0.708\tiny^{+0.198}_{-0.264}$ &\bf 0.002 $\tiny^{-0.013}_{+0.015}$ &-0.039  \\ \cline{9-10}
& 0.111 &(48.5/48)& (34.3/46)& ($2.9\sigma$) & $6.876\tiny^{+0.119}_{-0.147}$ & $1.128\tiny^{+0.206}_{-0.168}$ & $2.403\tiny^{+0.290}_{-0.268}$ & 0.013 &\bf -0.029 $\tiny^{-0.018}_{+0.022}$  \\ \hline

G&0.011& 1.202 & 0.831 & $1.4\times10^{-4}$ & $7.012\tiny^{+0.106}_{-0.105}$ & $0.939\tiny^{+0.140}_{-0.118}$ & $2.059\tiny^{+0.113}_{-0.226}$ &\bf-0.007 $\tiny^{-0.015}_{+0.015}$ & -0.049  \\ \cline{9-10}
& 0.060 & (54/45)&(35.7/43)& ($3.0\sigma$) & $6.802\tiny^{+0.104}_{-0.125}$ & 0.125$\pm${\tiny 0.0} & $0.518\tiny^{+0.163}_{-0.181}$ &  0.023 & \bf -0.018 $\tiny^{-0.015}_{+0.018}$  \\ \hline

H&-0.102& 1.596 & 0.789 & $2.0\times10^{-7}$ & $7.722\tiny^{+0.099}_{-0.108}$ & $1.191\tiny^{+0.107}_{-0.099}$ & $2.747\tiny^{+0.259}_{-0.227}$ &\bf-0.108 $\tiny^{-0.014}_{+0.015}$ &-0.155 \\ \cline{9-10}
& 0.173 & (76.6/48)& (36.3/46)& ($3.9\sigma$) & $5.359\tiny^{+0.186}_{-0.212}$ & 0.544$\pm${\tiny 0.0} & $0.647\tiny^{+0.135}_{-0.164}$ &  0.230  & \bf 0.198 $\tiny^{-0.028}_{+0.032}$  \\ \hline

I$\bf ^e$ &-0.103 & 1.691 & 0.845 & $1.9\times10^{-7}$ & $7.731\tiny^{+0.115}_{-0.129}$ & $1.181\tiny^{+0.128}_{-0.119}$ & $2.594\tiny^{+0.308}_{-0.266}$ &\bf-0.110 $\tiny^{-0.016}_{+0.018}$ &-0.157 \\ \cline{9-10}
& 0.174 &(81.2/48) & (38.9/46)& ($4.0\sigma$) & $5.221\tiny^{+0.238}_{-0.262}$ & 0.615$\pm${\tiny 0.0} & $0.748\tiny^{+0.166}_{-0.173}$ & 0.250 & \bf 0.219 $\tiny^{-0.035}_{+0.039}$  \\ \hline

J&-0.103 &  3.768 & 1.693 & $1.5\times10^{-8}$ & $7.693\tiny^{+ 0.085}_{-0.070}$ & $1.073\tiny^{+0.065}_{-0.069}$ & $6.149\tiny^{+0.367}_{-0.388}$ &\bf-0.104 $\tiny^{-0.012}_{+0.010}$ &-0.151 \\ \cline{9-10}
& 0.174 &(180.9/48) &(76/45) & ($4.3\sigma$) & $5.333\tiny^{+0.146}_{-0.146}$ & $0.567\tiny^{+0.211}_{-0.167}$ & $1.487\tiny^{+0.556}_{-0.376}$ & 0.234 & \bf 0.202 $\tiny^{-0.022}_{+0.022}$   \\ \hline

K$\bf ^f$ &-0.103 & 3.570 & 1.628 & $2.1\times10^{-8}$ & $7.587 \tiny^{+0.054}_{-0.032}$ & $1.011\tiny^{+0.056}_{-0.055}$ & $5.731\tiny^{+0.282}_{-0.284}$ &\bf-0.089 $\tiny^{-0.008}_{+0.005}$ & -0.135 \\ \cline{9-10}
& 0.174 &(171.4/48) &(73.3/45)& ($4.2\sigma$) & $5.323 \tiny^{+0.132}_{-0.141}$ & $0.495\tiny^{+0.148}_{-0.132}$ & $1.103\tiny^{+0.305}_{-0.244}$ &  0.236 &\bf 0.204 $\tiny^{-0.020}_{+0.021}$  \\ \hline

L&-0.103 & 3.176 & 1.597 & $1.8\times10^{-7}$ & $7.523\tiny^{+0.065}_{-0.058}$ & $1.031\tiny^{+0.061}_{-0.061}$ & $5.804\tiny^{+0.293}_{-0.322}$ & \bf-0.080 $\tiny^{-0.009}_{+0.008}$ &-0.126 \\ \cline{9-10}
&0.175 &(152.5/48) &(72.9/45)& ($4.0\sigma$) & $5.352\tiny^{+0.136}_{-0.069}$ & $0.462\tiny^{+0.158}_{-0.154}$ & $1.010\tiny^{+0.336}_{-0.247}$ &  0.231 &\bf 0.199 $\tiny^{-0.020}_{+0.010}$  \\ \hline

M&-0.096 & 1.454 & 0.845 & $4.4\times10^{-6}$ & $7.570\tiny^{+0.114}_{-0.099}$ & $1.014\tiny^{+0.104}_{-0.104}$ & $6.202\tiny^{+0.569}_{-0.591}$ &\bf-0.087 $\tiny^{-0.016}_{+0.014}$ &-0.132 \\ \cline{9-10}
& 0.168 &(69.8/48) &(38/45)& ($3.5\sigma$) & $5.453\tiny^{+0.209}_{-0.207}$ & $0.439\tiny^{+0.234}_{-0.237}$ & $1.204\tiny^{+0.589}_{-0.446}$ & 0.217 &\bf 0.184 $\tiny^{-0.031}_{+0.031}$  \\ \hline

\end{tabular}

\noindent{\small $a)$ Error bars are at $90$\% confidence level;
$b)$ Total photons cm$^{-2}$ sec$^{-1}$ in the line; $c)$
the fitting of the lower energy component was uncertain; $d)$ Massive
radio flare observed (Safi-Harb \& Kotani, 2003); \  $e)$ TOO observation at $\psi\sim 0$ and $\phi\sim 0$; \ $f)$ TOO
observation at $\psi\sim 0$ and $\phi\sim 0.5$}\\
\end{table}

Figure 1 shows background-subtracted light curves for the 
observations G, H, I, K and L (marked). The error-bars
obtained from counts$^{1/2}$/binsize are included. We extracted
the light curves both from the GoodXenon and Standard2 mode data.  The binsize was chosen
to be $16$s. The panel of Observation G was already  
presented in Chakrabarti et al. (2003), where, the multi-wavelength campaign was
reported. It was especially mentioned that this observation showed a
very rapid change in counts (around $15$\%) in a matter of minutes.
The panels of Observations H and I show the light curves of Oct. 1st,
2003 and Oct. 2nd 2003 respectively, the second observation being
at the donour inferior conjunction. While these light curves also show similar
rapid variation, there appears to be an overall modulation of X-ray counts.
At least $25$\% of the flux modulation occurs in 6-7 minutes timescale. Given that 
the companion star is directly blocking the base of the jet,
and the X-rays received may be passing through the star's
atmosphere this could even be due to some kind of oscillation in the atmosphere of the star.
By analysing both the GoodXenon and Standard2 mode data, we  could not
detect any quasi-periodicity in the power density spectra (PDS). The panel of 
Observation K shows the light curve of March the 13th, 2004 when the 
companion was right behind the disk. The jet pointing towards the observer had 
the highest blue-shift and the base of the jet was totally exposed. The 
X-ray flux was more than twice as high compared to average flux of past observations.
The X-ray count was also much higher on this date and a variation of 
$\sim 30\%$ was observed in a matter of few minutes.     
Similar result persisted on the next day (Observation L of 14th of March, 2004; 
see, Fig. 1) when short time-scale variability was more prominent. In fact, the 
X-ray flux remained high even after one orbital period (Observation M).

In the next section, we analyze line emissions from the RXTE observations listed in Table 1.

\section{Spectral fits of RXTE observations}

Recently, Marshall et al. (2002), using Chandra observation, pointed out that
the lines emitted do show blue- and red-shifted components. Migliari et al. (2002) reported
Chandra observation of iron lines emitted from extended regions of the jets. However,
similar observations from RXTE have not been reported so far. 

The data reduction and analysis was performed with the software HEASOFT 5.1
consisting of FTOOLS 5.1 and XSPEC 11.1. We extracted energy spectra from
PCA {\bf Standard-2} data.  For each spectrum, we have followed the standard procedures 
to generate the background spectra and PCA detector response matrices. We performed fitting the spectra 
simultaneously with different combination of models such as TB, line contribution, disk-blackbody 
spectrum and the power-law spectrum modified by interstellar absorption (WABS model, Morrison \& McCammon, 1983).

During fitting, we find that the so-called `traditional model' are best fitted with a minimum reduced $\chi^2$ 
value. We find good fitting while keeping the hydrogen column density fixed at $2.4 \times 10^{22}$cm$^{-3}$ 
except on Oct. 1st-2nd, 2003, when $1.6 \times 10^{22}$cm$^{-3}$ was needed. We did not set any systematic error. 
We also found that reduced $\chi^2$ is smaller if two lines  are included instead of a single line.
We have included in the Table 2 the  normal $\chi^2$ values and the degrees of freedom (dof) as well
as the reduced $\chi^2$ when one (denoted by $\chi_1^2$ in Column 3) or two (denoted by $\chi_2^2$ in Column 4) 
lines are fitted. We find that $\chi_2^2$ is always smaller than $\chi_1^2$.

In Fig. 2(a-c), we present raw spectra of the observations G, I, and K
showing a distinct bump at around $6-7$keV indicating the presence of Fe lines. 
On them, we superpose our folded spectra. While in (a), the precession/orbital phases were
generic (see Table 1), in (b), the phases were very close to zero when the companion blocked the
central compact object and in (c), the precession phase is close to zero but the orbital phase 
is close to $0.5$. 

\begin {figure}
\vbox{
\centerline{
\psfig{figure=f2a.ps,height=5.0truecm,width=5.0truecm,angle=270}
\hskip 0.0cm
\psfig{figure=f2b.ps,height=5.0truecm,width=5.0truecm,angle=270}
\hskip 0.0cm
\psfig{figure=f2c.ps,height=5.0truecm,width=5.0truecm,angle=270}}}
\vspace{1.0cm}
\noindent{\small {\bf Fig. 2a-c:} Raw X-ray spectra with the folded model of SS 433 on (a) obs. G on 27th September 2002,
(b) obs. I on 2nd of October, 2003 (inferior conjunction) and (c) obs. K on 
13th of March, 2004 (superior conjunction). Along the X-axis is energy (in keV) and along Y-axis
is the  normalized photon counts (sec$^{-1}$keV$^{-1}$). Precessional ($\psi$) and orbital 
($\phi$) phases are marked.}
\end{figure}

\begin {figure}
\vbox{
\centerline{
\psfig{figure=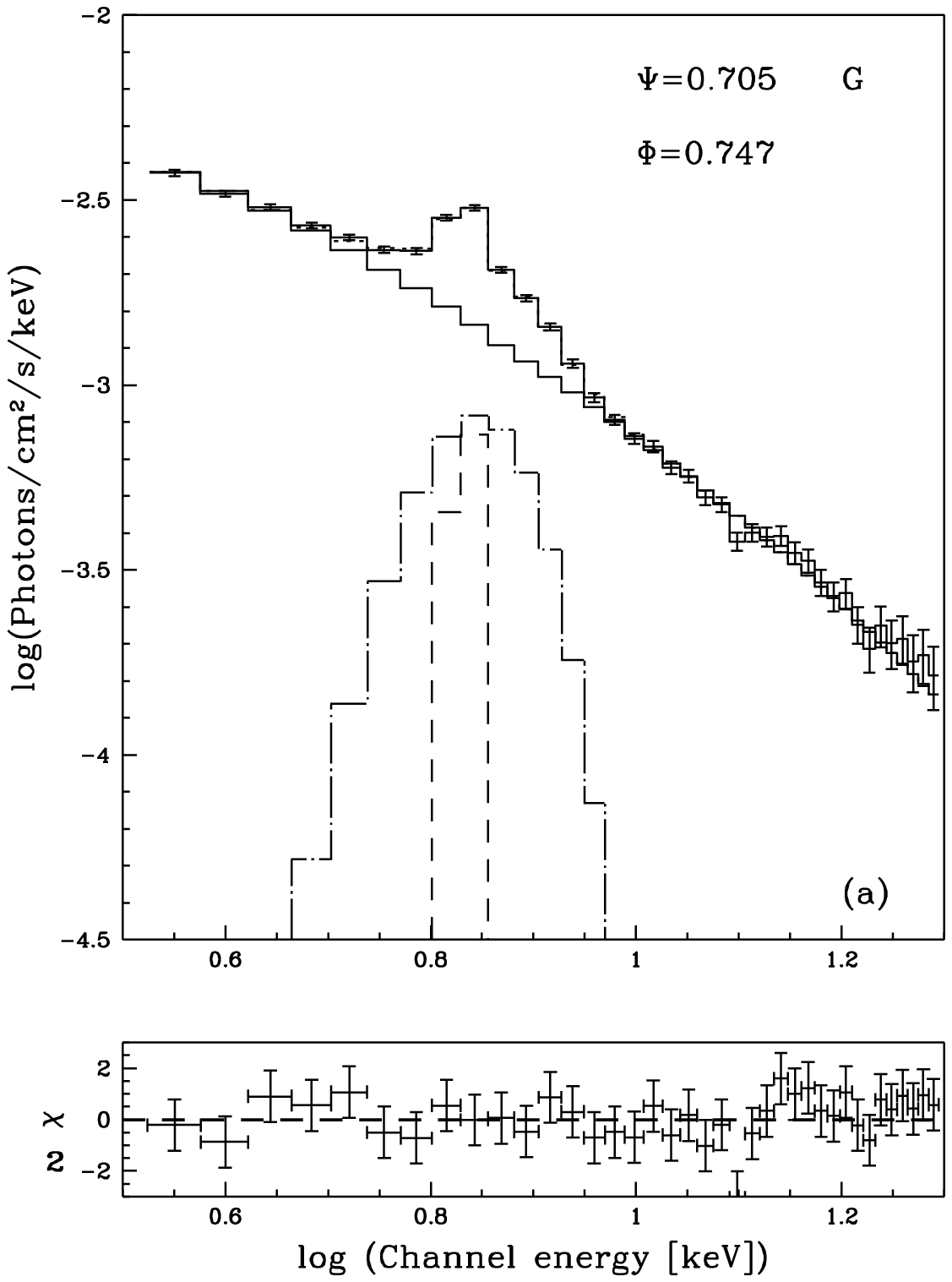,height=8truecm,width=8truecm}
\hskip -2.0cm
\psfig{figure=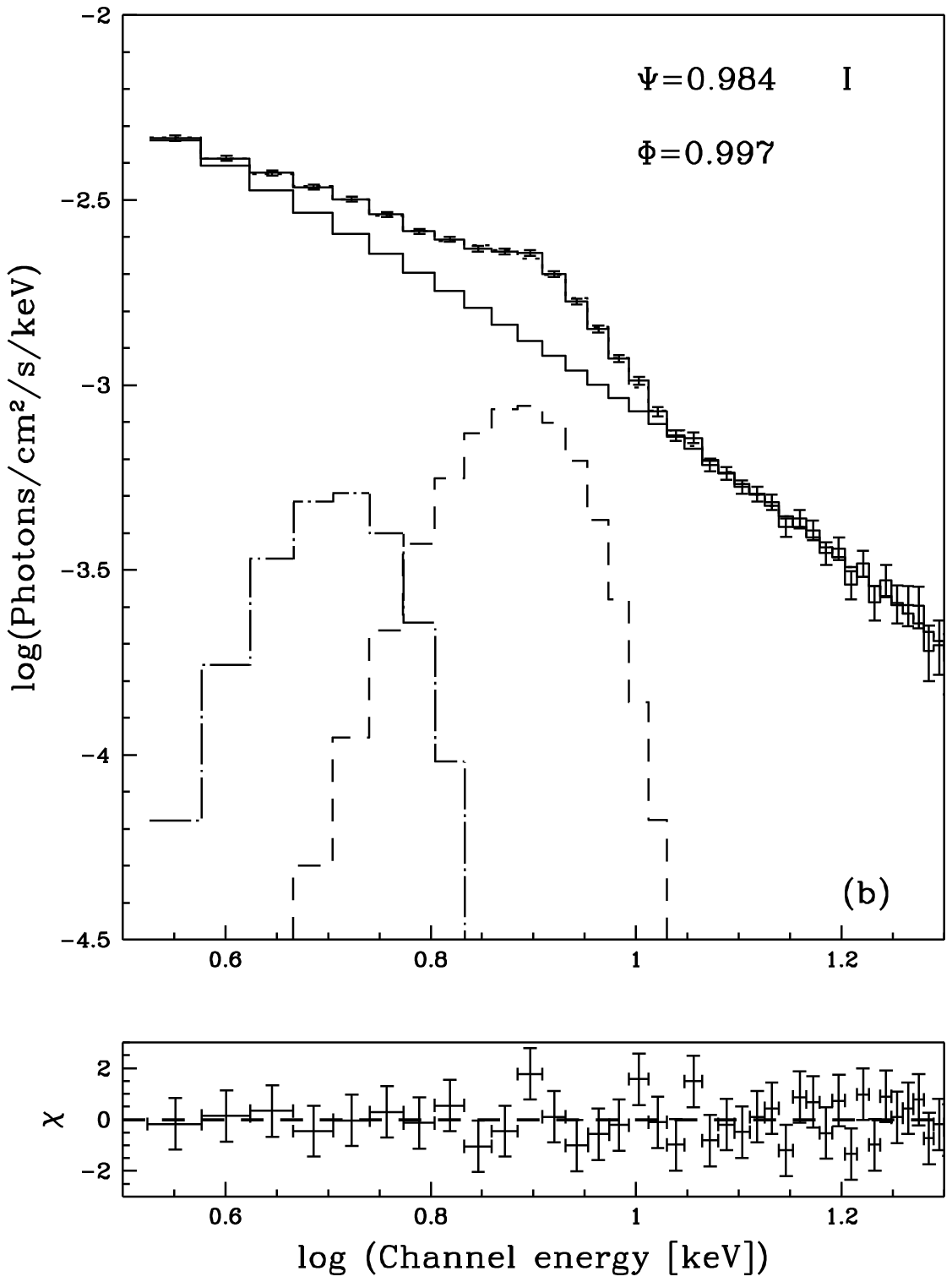,height=8truecm,width=8truecm}
\hskip -2.0cm
\psfig{figure=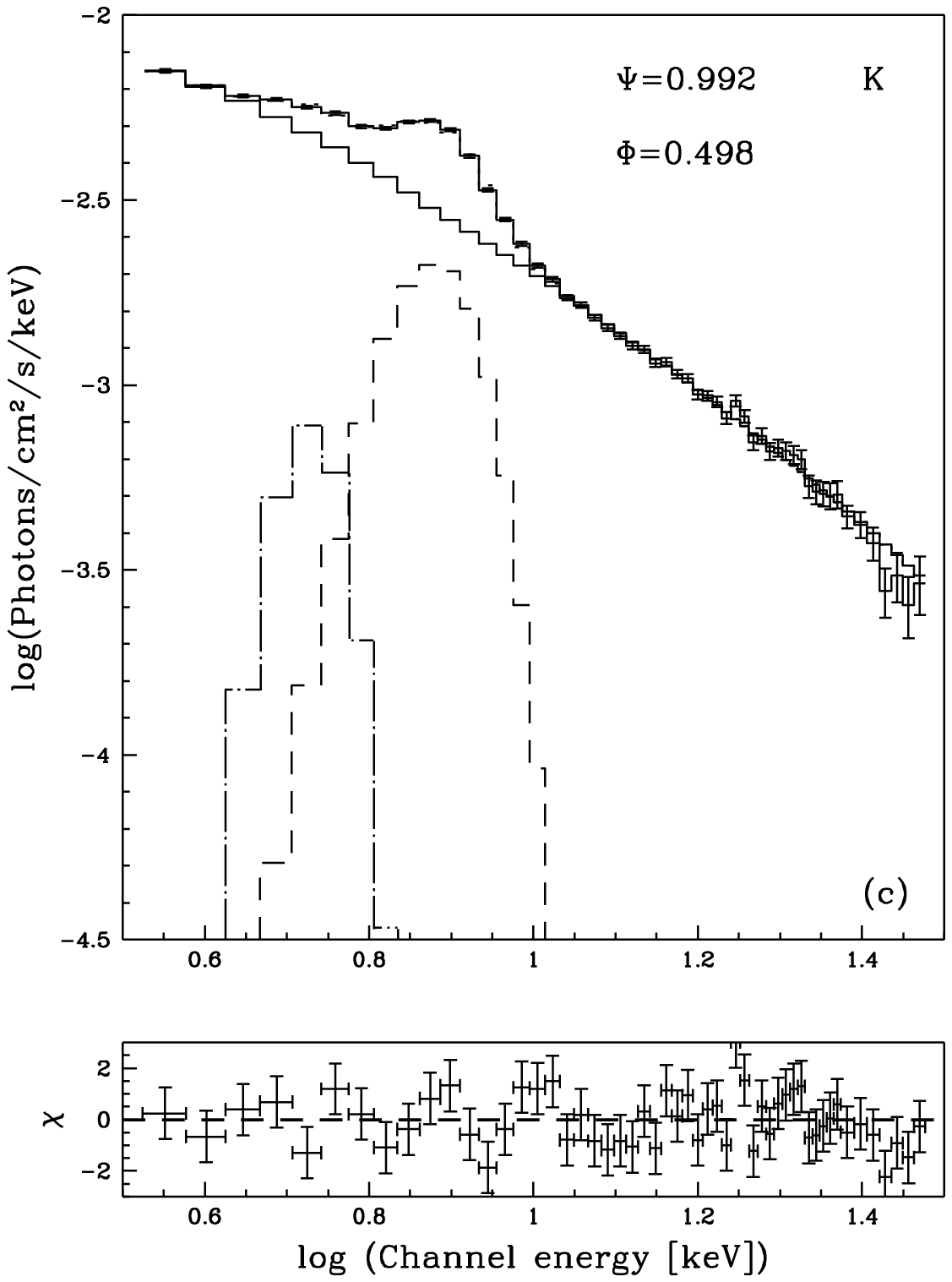,height=8truecm,width=8truecm}}}
\noindent{\small {\bf Fig. 3a-c:} Unfolded X-ray spectrum of SS 433 with different model components
for the same observations presented in Figs. 2(a-c). 
The model components are: thermal bremsstrahlung (solid histogram) and two Fe lines 
(dashed and long dash-dotted curves). In lower panels, residuals have also been shown. }
\end{figure}

For all observation, for the best fitting, two additional 
line features were tried out on the top of TB fit. 
In Fig. 3(a-c), we present the unfolded  spectra of the same observations, i.e., G (Fig. 3a), I  (Fig. 3b) 
and K (Fig. 3c) respectively with different model components.  The requirement of two line fitting in the 
spectrum is tested using the F-statistics with the {\it ftest} task within XSPEC. $P_{F-stat}$
is the F-statistic probability for the addition of the 2nd Gaussian line to the 
same model with a single line. This is given in Column 5 of Table 2. The significance
is given under this $P$ value and is denoted in parenthesis. Based on the significance value we find that
all the observations favour the two-line model except Obs. A and Obs. E as these fits are
significant only at $2.1 \sigma$ and $2.6 \sigma$ levels respectively. 
Indeed, we observe that on MJD 52234 (Observation E), the X-ray flux is very low. 
Given that it takes about a day for the jet 
to arrive from the X-ray emitting region to the radio emitting region, it is possible that the 
massive radio flare on MJD 52335 (Observation F. See, Safi-Harb \& Kotani 2003) actually was ejected on 
the previous day.  Similar behaviour of anti-correlation between Radio and X-ray 
fluxes has been reported by Mirabel \& Rodriguez (1999). 
During the final fit of the observations B, C, D, F and G, we froze the narrower line width.
In our fit, there was no signature of any soft-X-ray bump characteristics of a
Keplerian disk. As has been pointed out earlier (Chakrabarti et al. 2002), 
there is evidence that the flow is actually from a wind accretion
and thus possibly sub-Keplerian. This is also required for the production of the
observed jets (Chakrabarti 1999). The line energies (in keV) we obtained are given in Column 6. The residuals
given in the lower panels of Figs. 3(a-c) indicate that the fits are satisfactory.
Note that in Observations J-K-L, there are significant contributions from high energy 
($\sim 30keV$) photons and the flux is also much higher (see, Table 1). This behaviour persisted
even after one orbital period (Observation M). Columns 7 and 8 give the line width (keV) 
and line strength (in units of photons cm$^{-2}$ s$^{-1}$) respectively for each line. 
It has been noted earlier (Kotani et al. 1996, Marshall et al. 2002, Namiki et al. 2003) 
from the ASCA and Chandra observations that
the FeXXV lines are on an average 2-3 times (or more) stronger compared to FeXXVI lines.
In our RXTE observations C, D and F (when $\psi \sim 0.7$), we find roughly a similar result
when we identify both the lines to come from the approaching jet components. 
However, in observations B and G, where the lines were similarly identified as above,
the FeXXV line is found to be weaker compared to the FeXXVI line.
In observations H-M the jets are pointing towards the observer ($\psi \sim 0$)
and we find that two lines are roughly agreeing with the prediction of the kinematic model provided 
the line with higher energy (brighter component) is identified with 
FeXXVI of the approaching jet and the lower energy (dimmer component) is identified 
with FeXXV of the receding jet. In Observation E, we could fit with only one line, possibly
because of certain disturbances of the inner disk and the jet one day ahead of the
massive radio flare reported by Safi-Harb \& Kotani (2003).  

It is pertinent to ask whether one could have fitted the spectra with one Fe line and multiple TB
components instead of the way we fitted so far (single TB with two Fe lines). 
We have tried this for the thin Fe line occurring at low energy and found that the fit is deteriorated. For instance,
for Observation K, we replaced the Fe line of $5.323$keV by the TB of plasma electron temperature $0.5$keV
and $0.8$keV respectively (These temperatures were chosen in a way that the low energy
cut-off is at around $6$ keV.) and we found the reduced $\chi^2$ to be $3.586$ and $3.862$ respectively. 
However, with two Fe lines, the reduced $\chi^2$ is $1.628$ (see, Table 2). 
We therefore do not find evidence for multi-temperature bremsstrahlung components 
in our analysis. 

In Fig. 4(a-b), we drew contour plots of $\Delta \chi^2$ for Obs. C in the (a) line width
($\sigma$) vs. line energy ($E_{obs}$) plane and (b) line width ($\sigma$) vs. line flux ($F_{line}$)
plane to show the correlations.  Similarly, in Fig. 5(a-b), we drew the contour plots of  $\Delta \chi^2$ 
for Obs. K in the line width ($\sigma$) vs. line energy ($E_{obs}$) plane for the (a) broad line and (b) narrow line. 
The contours are of $68$\%, $90$\% and $99$\% confidence level.  It can be seen that the lines
are resolved at the 90\% confidence level.

While comparing with the absolute line strength with previous observations, we note that
Chandra (Marshall et al. 2002) obtained a red-jet FeXXV line (1s2p - 1s$^2$) flux of $\sim 0.13$ 
in the same unit chosen in Table 2, while we obtained the value of $\sim 1.2$ (Observation M), both observations
being at a similar precessional phase of $\psi \sim 0.06$ if the same ephemeris (Goranskii et al. 1998) 
were used. This high value may be because (a) the SS 433 was intrinsically brighter in X-ray in our observation and
(b) blending of lines which RXTE was unable to resolve especially there could be always a
significant contribution from the neutral Fe line emitted from regions 
at rest in the observed frame (Kotani et al. 1996, Marshall et al. 2002).

\begin {figure}
\vbox{
\vskip 1.0cm
\centerline{
\psfig{figure=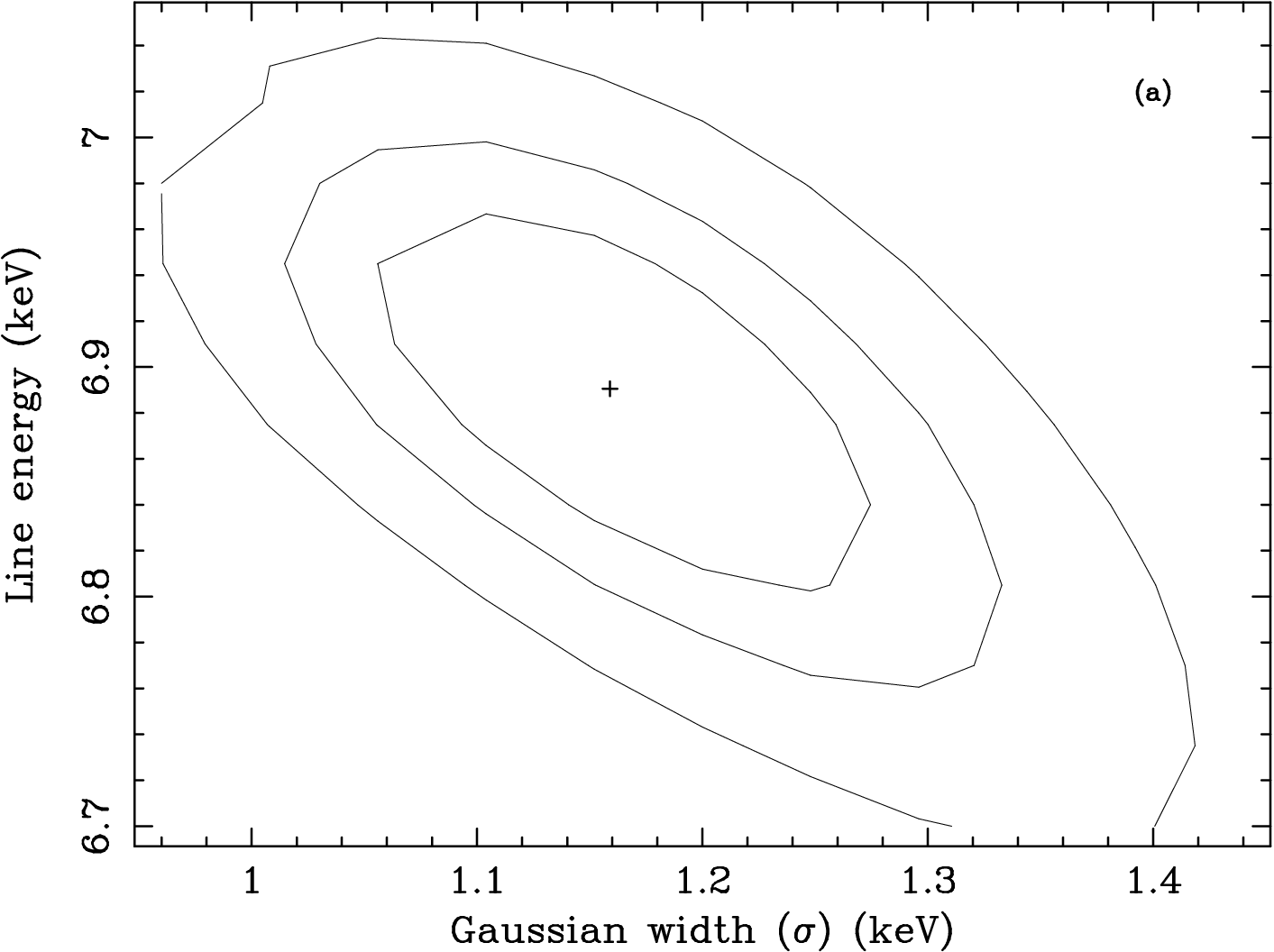,height=5truecm,width=7truecm,angle=270}
\hskip 0.5cm
\psfig{figure=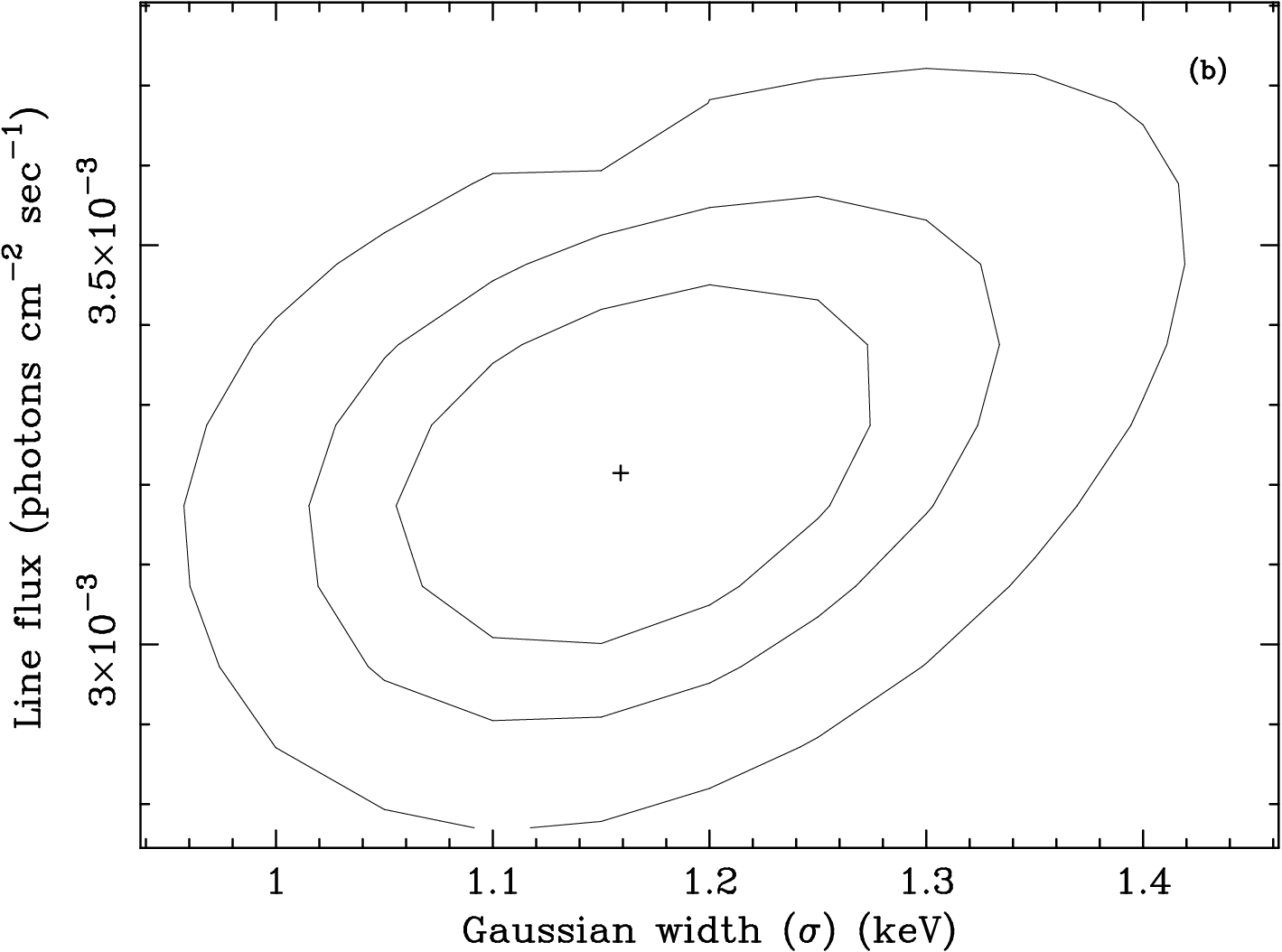,height=5truecm,width=7truecm,angle=270}}}
\vspace{1.0cm}
\noindent{\small {\bf Fig. 4a-b:} 
Two parameter confidence region of the broad FeXXV (a) line energy vs. Gaussian width
and (b) line flux vs. Gaussian width from spectral fitting to the RXTE/PCA data of 
Obs. C.  Contours correspond to 68\%, 90\% and 99\% confidence.}
\end{figure}

\begin {figure}
\vbox{
\vskip 1.0cm
\centerline{
\psfig{figure=f5a.ps,height=5truecm,width=7truecm,angle=270}
\hskip 0.5cm
\psfig{figure=f5b.ps,height=5truecm,width=7truecm,angle=270}}}
\vspace{1.0cm}
\noindent{\small {\bf Fig. 5a-b:} 
Two parameter confidence region of the (a) broad FeXXVI line energy vs. Gaussian width
and (b) narrow FeXXV line energy vs. Gaussian width from the spectral fitting to the RXTE/PCA
data of Obs. K. Contours correspond to 68\%, 90\% and 99\% confidence.}
\end{figure}


\begin {figure}
\vbox{
\vskip -2.0cm
\centerline{
\psfig{figure=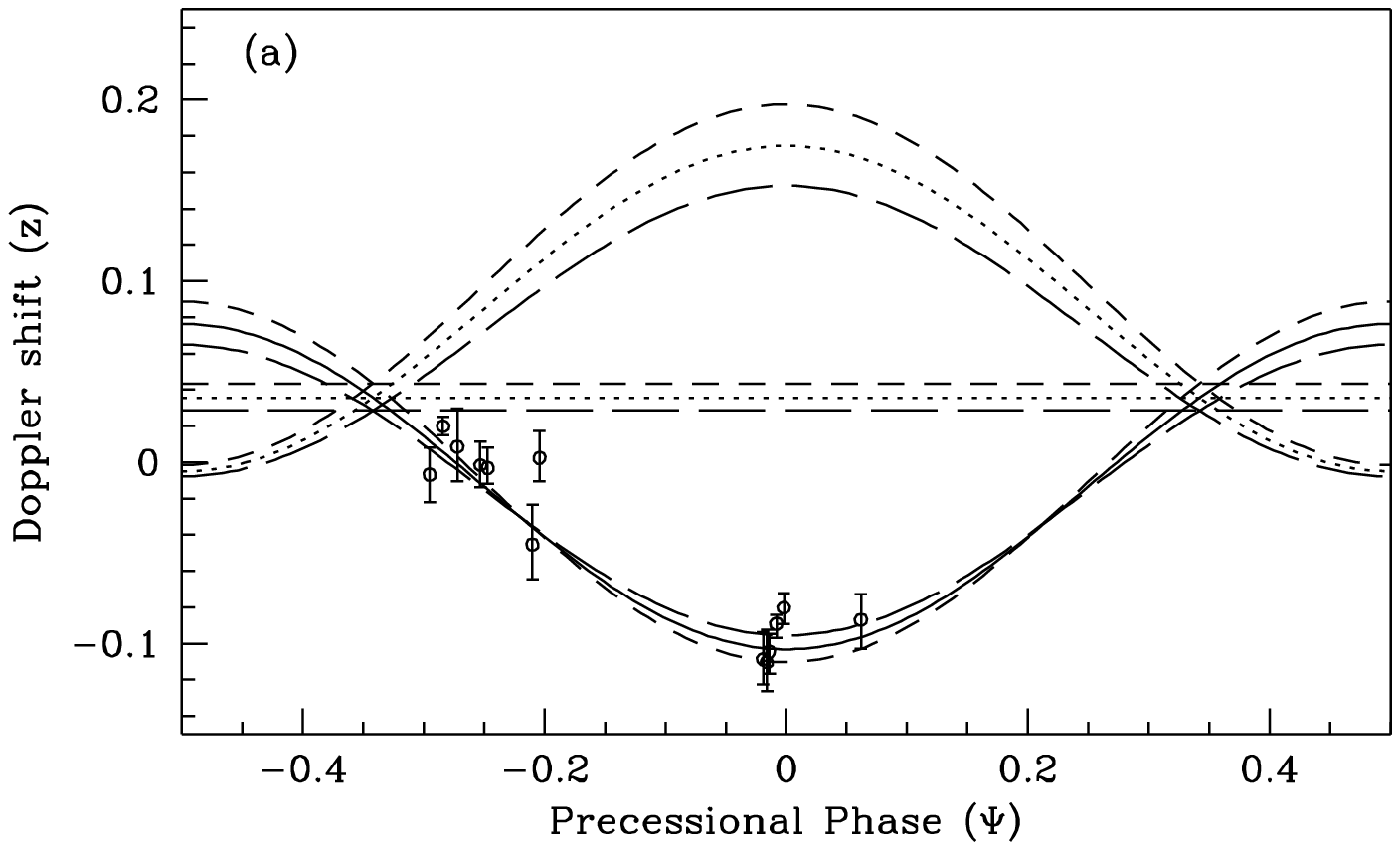,height=11truecm,width=11truecm}
\hskip -3.0cm
\psfig{figure=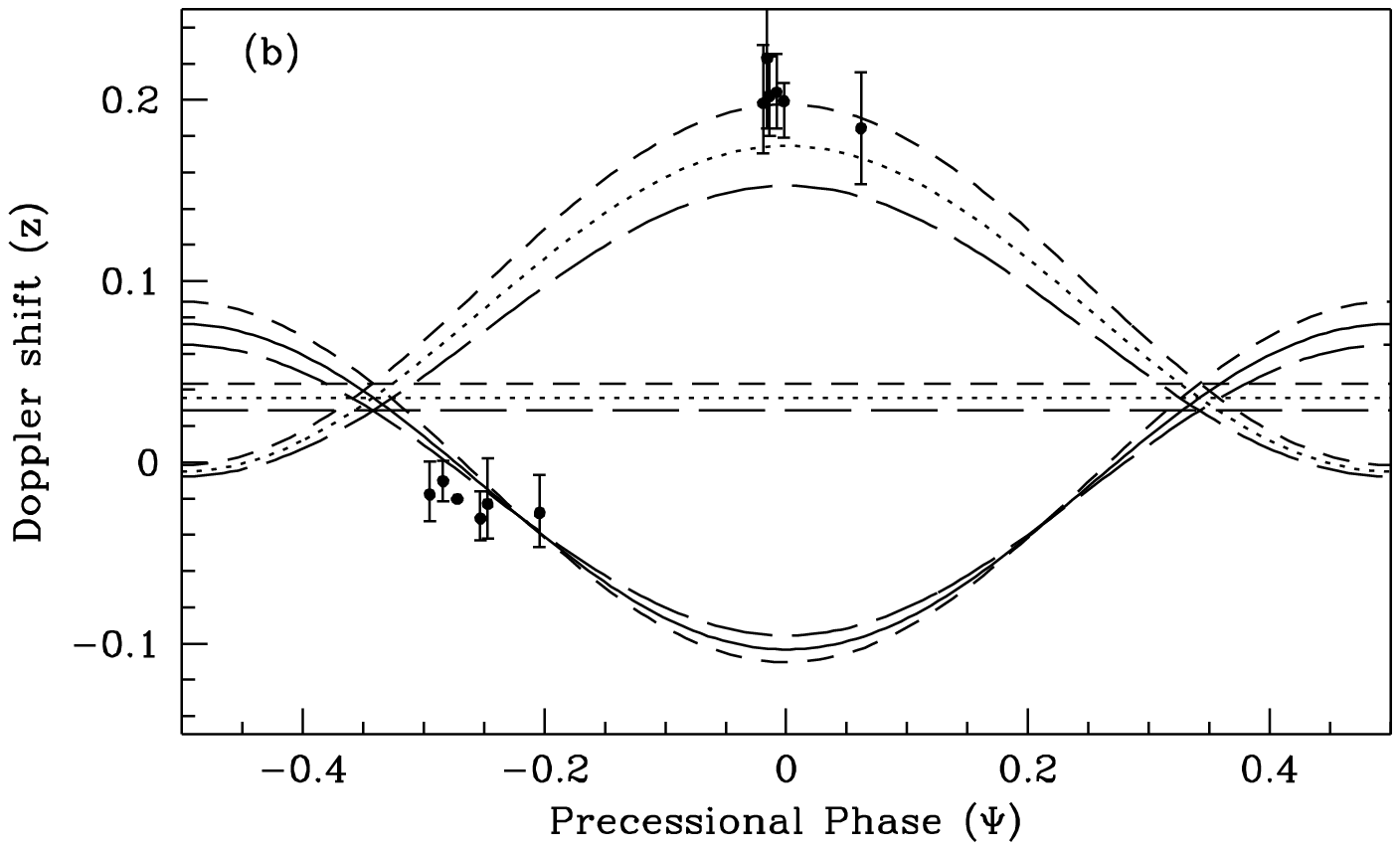,height=11truecm,width=11truecm}}}
\vspace{-2.0cm}
\noindent{\small {\bf Fig. 6a-b:} 
Observed Doppler shifts (y-axis) of (a) FeXXVI and (b) FeXXV lines from the approaching (solid curve)
and receding (dotted curve) jets super-imposed on the prediction from the kinematic model with 
standard parameters (see text) as a function of the precessional phase $\psi$.
Also shown in short-dashed and long-dashed curves the shifts obtained by respectively 
reducing and enhancing the jet velocity by ten percent. The horizontal curves
represent intrinsic red-shifts for the corresponding jet velocities. See text for
detailed criteria used in line-identifications. All the observations were plotted
against the same precessional phase for clarity of the plot of the lines.}
\end{figure}

In Columns 9-10 we present the computed  red- and blue-shift factors (z) of the observed
lines, had their origins been the FeXXVI (Ly $\alpha$ transition at $6.965$ keV) 
or the FeXXV line (1s2p - 1s$^2$ transition at $6.684$keV),
respectively. In bold faced letters we have highlighted the probable
identification of the lines. Generally speaking, the line with higher energy could be identified 
with the blue-shifted FeXXVI line quite satisfactorily (Fig. 6a). 
However, the line with lower energy could be fitted with  red-shifted FeXXV lines when $\psi \sim 0$ and  with
blue-shifted FeXXV line elsewhere (Fig. 6b). The error-bars (at 90\% confidence
level) drawn in Fig. 6(a-b) are given in Table 2. The data has been folded
with $162.15$d periodicity for convenience. Superimposed are the solid and dotted curves representing 
the Doppler shifts (Eq. 1) of the jet component pointing towards us and the 
component pointing away from us respectively. Though the poor resolution in RXTE/PCA detector
may be the main cause of the deviation of the fitted shifts from that of the kinematic model,
one can assume that the lines energies are correct in order to estimate the possible variation of 
jet velocity, if any, which may be responsible for this deviation, when other system
parameters are kept unchanged. We plot the short and long-dashed
curves in Figs. 6(a-b) for $v_j=0.286$ and $v_j=0.234$ respectively which are $10$\% away from
the velocity $v_j=0.2602$ of the standard model. The horizontal lines
correspond to the intrinsic red-shifts for these velocities. Since at $\psi\sim 0$, the blue
jet seems to have a large scattering of velocity, while the red jet seems to have a higher velocity,
we could not conclude with certainly if the jets have truly different velocity than that of the standard
kinematic model (e.g., Eikenberry et al. 2001), even though there are reports 
(e.g., Marshall et al. 2002) that the jet velocity could be higher close to the compact object.

We also examined if the lines could be fitted with NiXXVII (1s2p - 1s$^2$
transition at $7.788$ keV). However, except for the lower energy component of observation F,
none seems to be satisfactory. It may be noted that near $\psi\sim 0$, RXTE observation 
showed the evidence of both the components of the jet which has not been reported before.

\section{X-ray Flares in SS 433?}

During the recent TOO campaign of 12-14th March, 2004, the X-ray flux is found to be very high. 
This may indicate that SS 433 was undergoing some kind of weak X-ray `flare'. In order to search of past history of flares, we 
looked at the entire ASM rate profile. We plotted in
Fig. 7a,  the ASM rate since 7th Jan. 1996 till Nov. 26th, 2004
We plot  (solid) only those points when the count over the background rate 
is positive. This shows that there had been a few occurrences of `flaring' activities 
in the past, some weak and some strong. 
In Fig. 7b, we folded the ASM light curve 
around $T_{per}=368$ days of interval and taken the 
mean count along with standard deviation as the error bars to indicate that 
there are indeed some indication of periodicity. However, the folding 
looks similar even when $T_{per}$
is changed by $\sim 3$ days indicating that this might be an annual event 
related to ASM observation. While there are at least two other 
weaker peaks with this periodicity, they do not appear to be very
significant at this stage.

\begin {figure}
\vbox{
\vskip -3.0cm
\centerline{
\psfig{figure=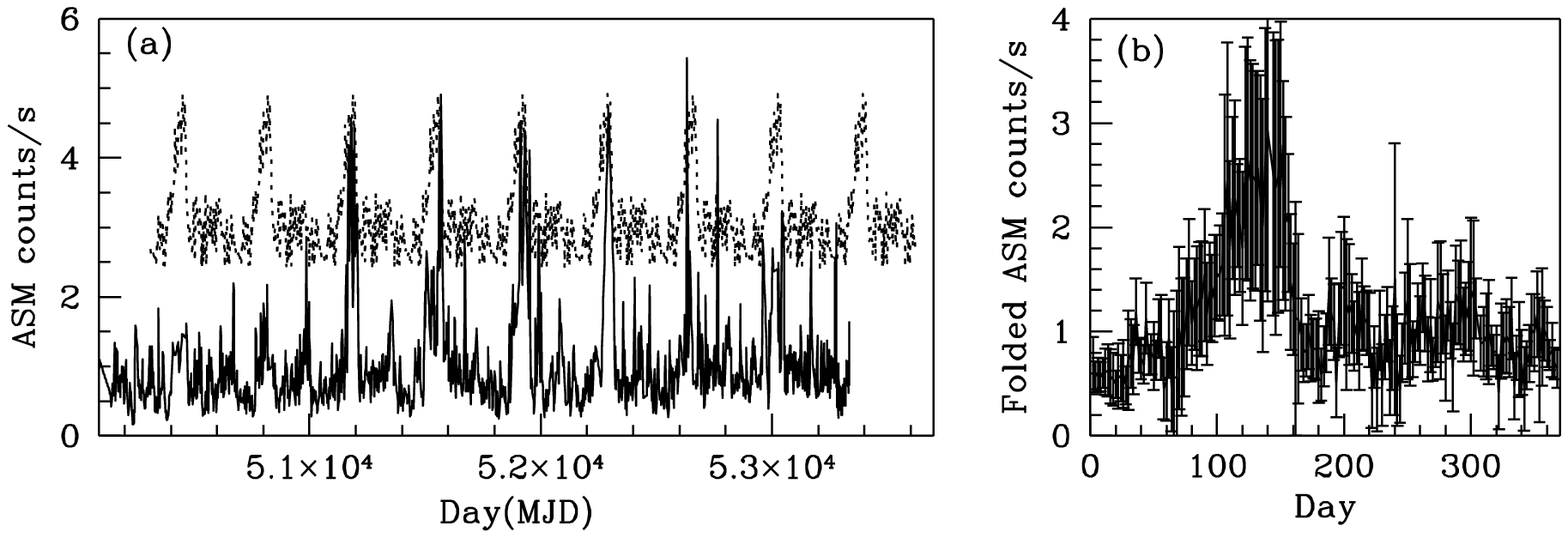,height=12truecm,width=16truecm}}}
\vspace{-3.0cm}
\noindent{\small {\bf Fig. 7a-b:}
(a) ASM counts/s of SS 433 since 7th Jan. 1996 till Nov. 26th, 2004. For clarity, we plotted only those
points when the count over the background rate is positive. We superpose (with a shift of $2$/s along y-axis)
the mean light-curve when folded around $368$ days to show the periodic nature of a 
`flare' which repeats after every $368$ days interval. This could be an artefact of the ASM obseration.
(b) Average ASM light curve obtained by  folding around $368$days showing 
the presence of the `flare'.}
\end{figure}

\section{Conclusion}

In the present paper, we analyzed a set of RXTE observations of SS 433 some of which were
triggered by us. We presented the results of opportune moments at inferior ($\phi\sim 0$) and superior
($\phi\sim 0.5$) conjunctions when the jets had the highest possible Doppler shifts ($\psi \sim 0$; 
Observations I and K respectively). We observed a considerable change in the emitted flux.
In particular, we observed a very high flux (more than twice the average flux seen in other days) 
in Observations J,K, L and M, the last one being taken one orbital period later than the previous one. We did 
not find any evidence of any short time-scale quasi-periodicity in the PCA light curves 
except perhaps a modulation at around $25\%$ level in time-scales of a few minutes, especially
when the companion blocked the base of the jet. It may be due to the oscillation of the atmosphere 
of the companion. During the superior conjunction the source was highly variable in a 
very short time-scale $<100$s. The spectra were best fitted with a model consisting of 
thermal bremsstrahlung component and emission lines. No signature of any Keplerian disk was found, 
possibly because of the obscuration of the inner part by matter gathered from the mass-loss of the 
companion. We generally found that two lines were required to fit the spectrum. The higher energy 
lines were generally identified with blue-shifted FeXXVI. The lower energy lines were generally 
identified with FeXXV blue-shifted or red-shifted depending on the precessional phases. 

Since RXTE/PCA detectors have poor energy resolution ($< 18$\% at $6$keV) compared to
ASCA and Chandra, we could not be sure if the deviation of the Doppler-shifts
from the value predicted by the kinematic model is solely due to deviation from the standard
jet-parameters, i.e., the velocity and orientation angles. It is possible that the jet velocity 
may be different (as inferred by  Marshall et al. 2002 using Chandra observation) or
there could be some effects due to the nodding motion which we did not include in our fitting.
Furthermore, the neutral Fe line at $z=0$ can also contribute to the flux (Kotani et al. 1996, Marshall et al. 2002)
We observed that on Oct.  2nd, 2003, the X-ray flux is lower (less than $10$\%) 
compared to that observed on Oct. 1st, 2003, only twelve hours earlier. 
This could be an indication that the X-ray source is progressively blocked by the
companion at the conjunction is reached. Similarly, on March 13th, 2004 (Observation K) the 
exposed jet was more than twice as bright and this trend continued even after one orbital period
(Observation M). These may also indicate that the base of the jet is perhaps the major source of X-rays in SS 433. 

From the ASM data we identified some `flaring' phenomena in SS 433. However, judging from the
fact that the periodicity is $\sim 1$yr, it is possible that this is an artefact of the observation.
There were smaller and weaker flares in ASM as well. In any case, we did observe very high flux in our
March, 2004 observation, so real flares do seem to take place in SS433.

\section*{Acknowledgments} We thank Jean Swank for giving us adequate RXTE time and fruitful
discussions. We also thank the anonymous referee for suggesting several improvements in data analysis.
The research of AN is partly supported by DST grant No. Grant No. SP/S2/K-15/2001.

{}

\end{document}